# A principled methodology for comparing relatedness measures for clustering publications

Ludo Waltman[1], Kevin W. Boyack[2], Giovanni Colavizza[3], and Nees Jan van Eck[1]

[1] Centre for Science and Technology Studies, Leiden University, The Netherlands
{waltmanlr, ecknjpvan}@cwts.leidenuniv.nl

[2] SciTech Strategies, Inc., Albuquerque, NM, USA
kboyack@mapofscience.com

[3] University of Amsterdam, The Netherlands
Centre for Science and Technology Studies, Leiden University, The Netherlands
g.colavizza@uva.nl

There are many different relatedness measures, based for instance on citation relations or textual similarity, that can be used to cluster scientific publications. We propose a principled methodology for evaluating the accuracy of clustering solutions obtained using these relatedness measures. We formally show that the proposed methodology has an important consistency property. The empirical analyses that we present are based on publications in the fields of cell biology, condensed matter physics, and economics. Using the BM25 text-based relatedness measure as evaluation criterion, we find that bibliographic coupling relations yield more accurate clustering solutions than direct citation relations and co-citation relations. The so-called extended direct citation approach performs similarly to or slightly better than bibliographic coupling in terms of the accuracy of the resulting clustering solutions. The other way around, using a citation-based relatedness measure as evaluation criterion, BM25 turns out to yield more accurate clustering solutions than other text-based relatedness measures.

## 1. Introduction

Clustering of scientific publications is an important problem in the field of bibliometrics. Bibliometricians have employed many different clustering techniques (e.g., Gläser, Scharnhorst, & Glänzel, 2017; Šubelj, Van Eck, & Waltman, 2016). In addition, they have used various different relatedness measures to cluster publications.

These relatedness measures are typically based on either citation relations (e.g., direct citation relations, bibliographic coupling relations, or co-citation relations) or textual similarity, or sometimes a combination of the two.

Which relatedness measure yields the most accurate clustering of publications? Two perspectives can be taken on this question. One perspective is that there is no absolute notion of accuracy (e.g., Gläser et al., 2017). Following this perspective, each relatedness measure yields clustering solutions that are accurate in their own right, and it is not meaningful to ask whether one clustering solution is more accurate than another one. For instance, different citation-based and text-based relatedness measures each emphasize different aspects of the way in which publications relate to each other, and the corresponding clustering solutions each provide a legitimate viewpoint on the organization of the scientific literature. The other perspective is that for some purposes it is useful, and perhaps even necessary, to assume the existence of an absolute notion of accuracy (e.g., Klavans & Boyack, 2017). When this perspective is taken, it is possible, at least in principle, to say that some relatedness measures yield more accurate clustering solutions than others.

We believe that both perspectives are useful. From a purely conceptual point of view, the first perspective is probably the more satisfactory one. However, from a more applied point of view, the second perspective is highly important. In many practical applications, users expect to be provided with a single clustering of publications. Users typically have some intuitive idea of accuracy and, based on this idea of accuracy, they expect the clustering provided to them to be as accurate as possible. In this paper, we take this applied viewpoint and we therefore focus on the second perspective.

Identifying the relatedness measure that yields the most accurate clustering of publications is challenging because of the lack of a ground truth. There is no perfect classification of publications that can be used to evaluate the accuracy of different clustering solutions. For instance, suppose we study the degree to which a clustering solution resembles an existing classification of publications (e.g., Haunschild, Schier, Marx, & Bornmann, 2018). The difficulty then is that it is not clear how discrepancies between the clustering solution and the existing classification should be interpreted. Such discrepancies could indicate shortcomings of the clustering solution, but they could equally well reflect problems of the existing classification.

As an alternative, the accuracy of clustering solutions can be evaluated by domain experts who assess the quality of different clustering solutions in a specific scientific domain (e.g., Šubelj et al., 2016). This approach has the difficulty that it is hard to find a sufficiently large number of experts who are willing to spend a considerable amount of time making a detailed assessment of the quality of different clustering solutions. Moreover, the knowledge of experts will often be restricted to relatively small domains, and it will be unclear to what extent the conclusions drawn by experts generalize to other domains.

In this paper, we take a large-scale data-driven approach to compare different relatedness measures based on which publications can be clustered. The basic idea is to cluster publications based on a number of different relatedness measures and to use another more or less independent relatedness measure as a benchmark for evaluating the accuracy of the clustering solutions. This approach has already been used extensively in a series of papers by Kevin Boyack, Dick Klavans, and colleagues. They compared different citation-based relatedness measures (Boyack & Klavans, 2010; Klavans & Boyack, 2017), including relatedness measures that take advantage of full-text data (Boyack, Small, & Klavans, 2013), as well as different text-based relatedness measures (Boyack et al., 2011). To evaluate the accuracy of clustering solutions, they used grant data, textual similarity (Boyack & Klavans, 2010; Boyack et al., 2011, 2013), and more recently also the reference lists of 'authoritative' publications, defined as publications with at least 100 references (Klavans & Boyack, 2017).[1]

Our aim in this paper is to introduce a principled methodology for performing analyses similar to the ones mentioned above. We restrict ourselves to the use of one specific clustering technique, namely the technique introduced in the bibliometric literature by Waltman and Van Eck (2012), but we allow the use of any measure of the relatedness of publications. For two relatedness measures $A$ and $B$, our proposed methodology offers a principled way to evaluate the accuracy of clustering solutions obtained using the two measures, where a third relatedness measure $C$ is used as the evaluation criterion. Unlike approaches taken in earlier papers, our methodology has an important consistency property.

---

[1] In a somewhat different context, the idea of evaluating two systems using a third more or less independent system as the evaluation criterion was explored by Li and Ruiz-Castillo (2013). These authors were interested in evaluating two classification systems for calculating field-normalized citation impact statistics, and they proposed to use a third independent classification system for performing the evaluation.

This paper is organized as follows. In Section 2, we introduce our methodology for evaluating the accuracy of clustering solutions obtained using different relatedness measures. In Section 3, we discuss the relatedness measures that we consider in our analyses. We report the results of the analyses in Section 4. We present comparisons of different citation-based and text-based relatedness measures that can be used to cluster publications. Our analyses are based on publications in the fields of cell biology, condensed matter physics, and economics. We summarize our conclusions in Section 5.

## 2. Methodology

To introduce our methodology for evaluating the accuracy of clustering solutions obtained using different relatedness measures, we first discuss the quality function that we use to cluster publications. We then explain how we evaluate the accuracy of a clustering solution, and we analyze the consistency of our evaluation framework. Finally, we discuss the importance of using an independent evaluation criterion.

### 2.1. Quality function for clustering publications

Consider a set of $N$ publications. Let $r_{ij}^X \geq 0$ denote the relatedness of publications $i$ and $j$ (with $i = 1, \ldots, N$ and $j = 1, \ldots, N$) based on relatedness measure $X$, and let $c_i^X \in \{1, 2, \ldots\}$ denote the cluster to which publication $i$ is assigned when publications are clustered based on relatedness measure $X$.

Publications are assigned to clusters by maximizing a quality function. We focus on the quality function of Waltman and Van Eck (2012). This quality function is given by

$$Q = \sum_{i,j} I(c_i^X = c_j^X)(r_{ij}^X - \gamma), \qquad (1)$$

where $I(c_i^X = c_j^X)$ equals 1 if $c_i^X = c_j^X$ and 0 otherwise and where $\gamma \geq 0$ denotes a so-called resolution parameter. The higher the value of this parameter, the larger the number of clusters that will be obtained. Hence, the resolution parameter $\gamma$ determines the granularity of the clustering. An appropriate value for this parameter can be chosen based on the specific purpose for which a clustering of publications is intended to be used. For some purposes it may be desirable to have a highly granular clustering, while for other purposes a less granular clustering may be preferable. Sjögårde and Ahlgren (2018, in press) proposed approaches for choosing the value of the resolution parameter $\gamma$ that allow clusters to be interpreted as research topics or specialties.

The quality function in (1) can also be written as

$$Q = \sum_{i,j} I(c_i^X = c_j^X) r_{ij}^X - \gamma \sum_k (s_k^X)^2, \qquad (2)$$

where $s_k^X$ denotes the number of publications assigned to cluster $k$, that is,

$$s_k^X = \sum_i I(c_i^X = k). \qquad (3)$$

We also refer to $s_k^X$ as the size of cluster $k$.

In the network science literature, the above quality function was proposed by Traag, Van Dooren, and Nesterov (2011), who referred to it as the constant Potts model. The quality function is closely related to the well-known modularity function introduced by Newman and Girvan (2004) and Newman (2004). However, as shown by Traag et al. (2011), it has the important advantage that it does not suffer from the so-called resolution limit problem (Fortunato & Barthélemy, 2007). Waltman and Van Eck (2012) introduced the above quality function in the bibliometric literature. In the field of bibliometrics, the quality function has been used by, among others, Boyack and Klavans (2014), Klavans and Boyack (2017), Perianes-Rodriguez and Ruiz-Castillo (2017), Ruiz-Castillo and Waltman (2015), Sjögårde and Ahlgren (2018, in press), Small, Boyack, and Klavans (2014), and Van Eck and Waltman (2014).

**2.2. Evaluating the accuracy of a clustering solution**

Suppose that we have three relatedness measures $A$, $B$, and $C$, and suppose also that we have used relatedness measures $A$ and $B$ to cluster a set of publications. Furthermore, suppose that we want to use relatedness measure $C$ to evaluate the accuracy of the clustering solutions obtained using relatedness measures $A$ and $B$. One way in which this could be done is by using relatedness measure $C$ to obtain a third clustering solution and by comparing the clustering solutions obtained using relatedness measures $A$ and $B$ with this third clustering solution. A large number of methods have been proposed for comparing clustering solutions (e.g., Fortunato, 2010). However, we do not take this approach. In order to have a consistent evaluation framework (see Subsection 2.3), we evaluate the accuracy of the clustering solutions obtained using relatedness measures $A$ and $B$ based directly on relatedness measure $C$, not based on a clustering solution obtained using this relatedness measure.

Let $A^{X|C}$ denote the accuracy of a clustering solution obtained using relatedness measure $X$ (with $X = A$ or $X = B$), where the accuracy is evaluated using relatedness measure $C$. We define $A^{X|C}$ as

$$A^{X|C} = \frac{1}{N}\sum_{i,j} I(c_i^X = c_j^X) r_{ij}^C. \qquad (4)$$

The clustering solution obtained using relatedness measure $A$ is considered to be more accurate than the clustering solution obtained using relatedness measure $B$ if $A^{A|C} > A^{B|C}$, and the other way around.

The above approach for evaluating the accuracy of a clustering solution favors less granular solutions over more granular ones. Of all possible clustering solutions, the least granular solution is the one in which all publications belong to the same cluster. According to (4), this least granular clustering solution always has the highest possible accuracy. There can be no other clustering solution with a higher accuracy. In order to perform meaningful comparisons, (4) should be used only for comparing clustering solutions that have the same granularity.

How do we determine whether two clustering solutions have the same granularity? We could require that both clustering solutions have been obtained using the same value for the resolution parameter $\gamma$. Alternatively, we could require that both clustering solutions consist of the same number of clusters. We do not take either of these approaches. Instead, we require that the sum of the squared cluster sizes is the same for two clustering solutions. In other words, two clustering solutions obtained using relatedness measures $A$ and $B$ have the same granularity if

$$\sum_k (s_k^A)^2 = \sum_l (s_l^B)^2. \qquad (5)$$

If (5) is satisfied, (4) can be used to compare in an unbiased way the clustering solutions obtained using relatedness measures $A$ and $B$. On the other hand, if (5) is not satisfied, a comparison based on (4) will be biased in favor of the less granular clustering solution. In practice, obtaining two clustering solutions that satisfy (5) typically will not be easy. For both clustering solutions, it may require a significant amount of trial and error with different values of the resolution parameter $\gamma$. In the end, it may turn out that (5) can be satisfied only approximately, not exactly. We will get back to this issue in Subsection 4.3.

A conceptual motivation for the evaluation framework introduced in this subsection is presented in Appendix A.1. This motivation is based on an analogy with the evaluation of the accuracy of different indicators that provide estimates of values drawn from a probability distribution.

**2.3. Consistency of the evaluation framework**

The choice of the accuracy measure defined in (4) and the granularity condition presented in (5) may seem quite arbitrary. However, provided that we use the quality function defined in (1), this choice has an important justification. Suppose that the accuracy of clustering solutions is evaluated using some relatedness measure $X$. Our choice of the accuracy measure in (4) and the granularity condition in (5) then guarantees that of all possible clustering solutions of a certain granularity the solution obtained using relatedness measure $X$ will be the most accurate one. In other words, it is guaranteed that $A^{X|X} \geq A^{Y|X}$ for any relatedness measure $Y$. This is a fundamental consistency property that we believe should be satisfied by any sound framework for evaluating the accuracy of clustering solutions obtained using different relatedness measures.

Suppose for instance that we have three clustering solutions, all of the same granularity, one solution obtained based on direct citation relations between publications, another one obtained based on bibliographic coupling relations, and a third one obtained based on co-citation relations. Suppose also that the accuracy of the clustering solutions is evaluated based on direct citation relations. It would then be a rather odd outcome if the clustering solution obtained based on bibliographic coupling or co-citation relations turned out to be more accurate than the solution obtained based on direct citation relations. In our evaluation framework, it is guaranteed that there can be no such inconsistent outcomes. When the accuracy of clustering solutions is evaluated based on direct citation relations, the clustering solution obtained based on direct citation relations will always be the most accurate one. We refer to Appendix B for a formal analysis of this important consistency property. The appendix also provides an example of an inconsistent evaluation framework.

**2.4. Independent evaluation criterion**

As already mentioned in Section 1, the approach that we take in this paper is to cluster publications based on a number of different relatedness measures and to use another more or less independent relatedness measure to evaluate the accuracy of the clustering solutions. Our idea is to consider different relatedness measures as different proxies of the same underlying notion of relatedness. This underlying notion of relatedness, which may be referred to as the 'true' relatedness of publications, cannot be directly observed. It can only be approximated. Given the notion of the true relatedness of publications, each relatedness measure provides both signal and noise. To the extent that a relatedness measure approximates the true relatedness of publications, it provides signal. For the rest, it provides noise. We consider two relatedness measures to be independent if their noise is uncorrelated. For instance, a citation-based and a text-based relatedness measure may be considered independent. They are both noisy, but in quite different ways. On the other hand, two citation-based relatedness measures may not be considered independent. Both relatedness measures can be expected to be affected by similar types of noise, for instance noise caused by the fact that the authors of a publication cite a specific reference while some other reference would have been more relevant.

In this paper, we use a text-based relatedness measure to evaluate the accuracy of different clustering solutions obtained using citation-based relatedness measures, and the other way around, we use a citation-based relatedness measure to evaluate the accuracy of different clustering solutions obtained using text-based relatedness measures. Importantly, we are not interested in evaluating citation-based clustering solutions using a citation-based relatedness measure or text-based clustering solutions using a text-based relatedness measure. Such evaluations are of little interest because the relatedness measure used for evaluation is not sufficiently independent of the relatedness measures being evaluated. For instance, when direct citation relations are used to evaluate the accuracy of different clustering solutions obtained using citation-based relatedness measures, the clustering solution obtained based on direct citation relations will be the most accurate one. The evaluation simply shows that the clustering solution obtained based on direct citation relations is best aligned with an evaluation criterion based on direct citation relations, which of course is not surprising. This illustrates the importance of using an independent evaluation criterion. The more the relatedness measure used for evaluation can be considered to be independent of the relatedness measures being evaluated, the more informative the evaluation will be.

In Appendix A.2, we provide a further demonstration of the importance of using an independent evaluation criterion.

## 3. Relatedness measures

We now discuss the relatedness measures that we consider in this paper. We first discuss relatedness measures based on citation relations, followed by relatedness measures based on textual similarity. We also discuss the so-called top $M$ relatedness approach as well as the idea of normalized relatedness measures.

### 3.1. Citation-based relatedness measures

Below we discuss a number of citation-based approaches for determining the pairwise relatedness for a set of $N$ publications. We use $c_{ij}$ to indicate whether publication $i$ cites publication $j$ ($c_{ij} = 1$) or not ($c_{ij} = 0$).

The relatedness of publications $i$ and $j$ based on direct citation relations is given by

$$r_{ij}^{\text{DC}} = \max(c_{ij}, c_{ji}). \tag{6}$$

Hence, $r_{ij}^{\text{DC}} = 1$ if publication $i$ cites publication $j$ or the other way around and $r_{ij}^{\text{DC}} = 0$ if neither publication cites the other.

The relatedness of publications $i$ and $j$ based on bibliographic coupling relations equals the number of common references in the two publications. This can be written as

$$r_{ij}^{\text{BC}} = \sum_k c_{ik} c_{jk}, \tag{7}$$

where the summation extends over all publications in the database that we use, not only over the $N$ publications for which we aim to determine their pairwise relatedness.

As is well known, co-citation can be seen as the opposite of bibliographic coupling. The relatedness of publications $i$ and $j$ based on co-citation relations equals the number of publications in which publications $i$ and $j$ are both cited. In mathematical terms,

$$r_{ij}^{\text{CC}} = \sum_k c_{ki} c_{kj}, \tag{8}$$

where the summation again extends over all publications in the database that we use.

The above approaches for determining the relatedness of publications may also be combined. This results in

$$r_{ij}^{\text{DC–BC–CC}} = \alpha r_{ij}^{\text{DC}} + r_{ij}^{\text{BC}} + r_{ij}^{\text{CC}}, \tag{9}$$

where $\alpha$ denotes a parameter that determines the weight of direct citation relations relative to bibliographic coupling and co-citation relations. A direct citation relation may be considered a stronger signal of the relatedness of two publications than a bibliographic coupling or co-citation relation (Waltman & Van Eck, 2012), and therefore one may want to give more weight to a direct citation relation than to the two other types of relations. This can be achieved by setting $\alpha$ to a value above 1. The idea of combining different types of citation-based relations is not new. This idea was also explored by Small (1997) and Persson (2010).

In addition to the above citation-based approaches for determining the relatedness of publications, we also consider a so-called extended direct citation approach. Like the ordinary direct citation approach, the extended direct citation approach takes into account only direct citation relations between publications. However, direct citation relations are considered not just within the set of $N$ focal publications but within an extended set of publications. In addition to the $N$ focal publications, the extended set of publications includes all publications in our database that have a direct citation relation with at least two focal publications. (Publications that have a direct citation relation with only one focal publication are not considered because they do not contribute to improving the clustering of the focal publications.) The technical details of the extended direct citation approach are somewhat complex. These details are discussed in Appendix C. We note that an approach similar to our extended direct citation approach was also used by Boyack and Klavans (2014) and Klavans and Boyack (2017).

**3.2. Text-based relatedness measures**

We consider two text-based approaches for determining the relatedness of publications. We use $o_{il}$ to denote the number of occurrences of term $l$ in publication $i$. To count the number of occurrences of a term in a publication, only the title and abstract of the publication are considered, not the full text. Part-of-speech tagging is applied to the title and abstract of the publication to identify nouns and adjectives. The part-of-speech tagging algorithm provided by the Apache OpenNLP 1.5.2 library is used. A term is defined as a sequence of nouns and adjectives, with the last word in the sequence being a noun. No distinction is made between singular and plural nouns, so *neural network* and *neural networks* are regarded as the same term. Furthermore, shorter terms embedded in longer terms are not counted. For instance, if a publication contains the term *artificial neural network*, this is counted as an occurrence of *artificial neural network* but not as an occurrence of *neural network* or *network*. Finally, no stop word list is used, so there are no terms that are excluded from being counted.

A straightforward text-based measure of the relatedness of publications $i$ and $j$ is given by

$$r_{ij}^{\text{CT}} = \sum_l \frac{o_{il} o_{jl}}{(\sum_k o_{kl})^\beta}. \tag{10}$$

We refer to this as relatedness based on common terms. The denominator in (10) aims to reduce the influence of frequently occurring terms. The parameter $\beta$ in the denominator determines the extent to which the influence of these terms is reduced. If $\beta = 0$, no reduction in the influence of frequently occurring terms takes place. On the other hand, if $\beta = 1$, the influence of frequently occurring terms is strongly reduced, following a so-called fractional counting approach (Perianes-Rodriguez, Waltman, & Van Eck, 2016).

Boyack et al. (2011) identified BM25 as one of the most accurate text-based relatedness measures for clustering publications. We therefore also include BM25 in our analysis. BM25 originates from the field of information retrieval, where it is used to determine the relevance of a document for a search query (Sparck Jones, Walker, & Robertson, 2000a, 2000b). Following Boyack et al. (2011), we use BM25 as a text-based measure of the relatedness of publications. The BM25 relatedness measure is defined as

$$r_{ij}^{\text{BM25}} = \sum_l I(o_{il} > 0) \, \text{IDF}_l \frac{o_{jl}(k_1+1)}{o_{jl}+k_1\left(1-b+b\frac{d_j}{\bar{d}}\right)}, \quad (11)$$

where $I(o_{il} > 0)$ equals 1 if $o_{il} > 0$ and 0 otherwise and where $d_j$ and $\bar{d}$ denote, respectively, the length of publication $j$ and the average length of all $N$ publications. We define the length of a publication as the total number of occurrences of terms in the publication. This results in

$$d_i = \sum_l o_{il} \quad \text{and} \quad \bar{d} = \frac{1}{N}\sum_i d_i. \quad (12)$$

$\text{IDF}_l$ in (11) denotes the inverse document frequency of term $l$, which we define as

$$\text{IDF}_l = \log \frac{N-n_l+0.5}{n_l+0.5}, \quad (13)$$

where $n_l$ denotes the number of publications in which term $l$ occurs, that is,

$$n_l = \sum_i I(o_{il} > 0). \quad (14)$$

The BM25 relatedness measure in (11) depends on the parameters $k_1$ and $b$. Following Boyack et al. (2011), we set these parameters to values of 2 and 0.75, respectively. Unlike all other relatedness measures that we consider in this paper, the BM25 relatedness measure is not symmetrical. In other words, $r_{ij}^{\text{BM25}}$ does not need to be equal to $r_{ji}^{\text{BM25}}$.

**3.3. Top *M* relatedness approach**

Our interest focuses on large-scale clustering analyses that may involve hundreds of thousands or even millions of publications. These analyses impose significant challenges in terms of computing time and memory requirements. In particular, in these analyses, it may not be feasible to store all non-zero relatedness values in the main memory of the computer that is used.

To deal with this problem, we use the top *M* relatedness approach. This approach is quite similar to the idea of similarity filtering typically used by Kevin Boyack and Dick Klavans (e.g., Boyack & Klavans, 2010; Boyack et al., 2011). In the top *M* relatedness approach, only the top *M* strongest relations per publication are kept. (Ties are broken randomly.) The remaining relations are discarded. We use $\tilde{r}_{ij}^X$ to denote the relatedness of publications $i$ and $j$ based on relatedness measure *X* after discarding relations that are not in the top *M* per publication. This means that $\tilde{r}_{ij}^X = r_{ij}^X$ if publication $j$ is among the *M* publications that are most strongly related to publication $i$ and that $\tilde{r}_{ij}^X = 0$ otherwise. Relatedness of a publication with itself is ignored. Hence, $\tilde{r}_{ij}^X = 0$ if $i = j$. In general, $\tilde{r}_{ij}^X$ will not be symmetrical.

In most of the analyses presented in this paper, we use a value of 20 for *M*, although we also explore alternative values. We apply the top *M* relatedness approach to all our relatedness measures except for the measures based on (extended) direct citation relations. As pointed out by Waltman and Van Eck (2012), the use of direct citation relations has the advantage of requiring only a relatively limited amount of computer memory, and therefore there is no need to use the top *M* relatedness approach when working with direct citation relations. Applying the top *M* relatedness approach in the case of direct citation relations would also be problematic because all relations are equally strong, making it difficult to decide which relations to keep and which ones to discard. Hence, in the case of direct citation relations, we simply have $\tilde{r}_{ij}^{\text{DC}} = r_{ij}^{\text{DC}}$ for all publications $i$ and $j$.

**3.4. Normalization of relatedness measures**

We also normalize all relatedness measures. The normalized relatedness of publication $i$ with publication $j$ equals the relatedness of publication $i$ with publication $j$ divided by the total relatedness of publication $i$ with all publications. Hence, the normalized relatedness of publication $i$ with publication $j$ based on relatedness measure $X$ is given by

$$\hat{r}_{ij}^X = \frac{\tilde{r}_{ij}^X}{\sum_k \tilde{r}_{ik}^X}. \qquad (15)$$

This normalization was also used by Waltman and Van Eck (2012). The idea of the normalization is that relatedness values of publications in different fields of science should be of the same order of magnitude, so that clusters in different fields will be of similar size. Without the normalization, citation-based relatedness values for instance can be expected to be much higher in the life sciences than in the social sciences. In a clustering analysis that involves both publications in the life sciences and publications in the social sciences, this would result in life science clusters being systematically larger than social science clusters. The normalization in (15) can be used to correct for such differences between fields. The normalization also has the advantage that, regardless of the choice of a relatedness measure, a specific value of the resolution parameter $\gamma$ will always yield clustering solutions that have approximately the same granularity.

All results presented in the next section are based on normalized relatedness measures.

## 4. Results

We start the discussion of the results of our analyses by explaining the data collection and the way in which publications were clustered. We then introduce the idea of granularity-accuracy plots. Next, we present a comparison of different citation-based relatedness measures that can be used to cluster publications. This is followed by a comparison of different text-based relatedness measures.

**4.1. Data collection**

Data was collected from the Web of Science database. We used the in-house version of the Web of Science database available at the Centre for Science and Technology Studies at Leiden University. This version of the database includes the Science Citation Index Expanded, the Social Sciences Citation Index, and the Arts & Humanities Citation Index.

Like in our earlier work (e.g., Klavans & Boyack, 2017; Waltman & Van Eck, 2012), our final interest is in clustering all publications available in the database that we use, without restricting ourselves to certain fields of science. However, to keep the analyses presented in this paper manageable, we restricted ourselves to three specific fields. We selected all publications of the document types *article* and *review* that appeared in the period 2007–2016 in journals belonging to the Web of Science subject categories *Cell biology*, *Physics, condensed matter*, and *Economics*. Our aim was to cover three broad scientific domains, namely the life sciences, the physical sciences, and the social sciences. The subject categories *Cell biology*, *Physics, condensed matter*, and *Economics* were chosen because they cover these three domains and because they are relatively large in terms of the number of publications they include. The number of publications is 252,954 in cell biology, 272,935 in condensed matter physics, and 172,690 in economics.

The relatedness measures discussed in Section 3 were calculated for the selected publications. Two comments need to be made. First, in determining bibliographic coupling relations between publications, only common references to publications indexed in our Web of Science database were considered. This database includes publications starting from 1980. Common references to non-indexed publications (e.g., books, conference proceedings publications, and PhD theses) were not taken into account. Non-indexed publications were not considered in the extended direct citation approach either. Second, when we collected the data in Spring 2017, our database included a limited number of publications from 2017. These publications were not used in determining co-citation relations between publications. They also were not considered in the extended direct citation approach.

Table 1 reports for each of the three fields of science that we analyze and for each of the relatedness measures that we consider the average number of relations per publication and the percentage of publications that have no relations at all. The average number of relations per publication was calculated after applying the top *M* relatedness approach (except for DC and EDC; see Subsection 3.3). Table 1 shows that in the case of DC and especially CC a quite high percentage of the publications have no relations. This can be expected to have a negative effect on the accuracy of clustering solutions obtained using these relatedness measures, since publications without relations cannot be properly clustered.

Table 1. The average number of relations per publication (ANR) and the percentage of publications without relations (PWR) for different fields of science and different citation-based and text-based relatedness measures.

|  | Cell biology | | Condensed matter physics | | Economics | |
|---|---|---|---|---|---|---|
|  | ANR | PWR | ANR | PWR | ANR | PWR |
| DC | 11.3 | 8.5% | 7.5 | 12.3% | 8.0 | 11.0% |
| BC | 32.4 | 0.5% | 31.3 | 1.0% | 30.6 | 4.3% |
| CC | 25.7 | 13.5% | 19.6 | 20.0% | 16.9 | 30.7% |
| DC-BC-CC ($\alpha = 1$) | 32.3 | 0.4% | 31.3 | 0.7% | 30.9 | 2.7% |
| DC-BC-CC ($\alpha = 5$) | 31.6 | 0.4% | 30.5 | 0.7% | 29.8 | 2.7% |
| EDC | 69.0 | 0.3% | 39.5 | 0.7% | 24.2 | 2.6% |
| BM25 | 31.7 | 0.0% | 32.0 | 0.3% | 32.1 | 0.2% |
| CT ($\beta = 0.0$) | 38.1 | 0.0% | 38.6 | 0.3% | 38.5 | 0.2% |
| CT ($\beta = 0.5$) | 31.0 | 0.0% | 29.6 | 0.3% | 30.3 | 0.2% |
| CT ($\beta = 1.0$) | 26.3 | 0.0% | 26.8 | 0.3% | 27.0 | 0.2% |

**4.2. Clustering of publications**

For each of our three fields (i.e., cell biology, condensed matter physics, and economics), the selected publications were clustered based on each of our relatedness measures. Clustering was performed by maximizing the quality function presented in (1). To maximize the quality function, we used an iterative variant (Waltman & Van Eck, 2013) of the well-known Louvain algorithm (Blondel, Guillaume, Lambiotte, & Lefebvre, 2008). Five iterations of the algorithm were performed. In addition, to speed up the algorithm, we employed ideas similar to the pruning idea of Ozaki, Tezuka, and Inaba (2016) and the prioritization idea of Bae, Halperin, West, Rosvall, and Howe (2017). Our algorithm is a predecessor of the recently introduced Leiden algorithm (Traag, Waltman, & Van Eck, 2019), which was not yet available when we carried out our analyses. In general, our algorithm will not be able to find the global maximum of the quality function, but it can be expected to get close to the global maximum.

Different levels of granularity were considered. For each relatedness measure, we obtained ten clustering solutions, each of them for a different value of the resolution parameter $\gamma$. The following values of $\gamma$ were used: 0.00001, 0.00002, 0.00005, 0.0001, 0.0002, 0.0005, 0.001, 0.002, 0.005, and 0.01. Because of the normalization discussed in Subsection 3.4, the same values of $\gamma$ could be used for all relatedness measures. Without the normalization, different values of $\gamma$ would need to be used for each of the relatedness measures.

**4.3. Granularity-accuracy plots**

A difficulty of the evaluation framework presented in Subsection 2.2 is the requirement that the clustering solutions being compared have exactly the same granularity. This requirement, which is formalized in the condition in (5), is hard to meet in practice. Clustering solutions obtained using different relatedness measures but the same value of the resolution parameter $\gamma$ will approximately satisfy (5), but the condition normally will not be satisfied exactly.

To deal with this problem, we propose a graphical approach based on the idea of granularity-accuracy (GA) plots. Using a GA plot, relatedness measures can be compared despite differences in granularity between clustering solutions. The horizontal axis in a GA plot represents the granularity of a clustering solution. We define the granularity of a clustering solution obtained using relatedness measure $X$ as

$$\frac{N}{\sum_k (s_k^X)^2}. \qquad (16)$$

Two clustering solutions that have the same granularity according to (16) indeed satisfy the condition in (5). The vertical axis in a GA plot represents the accuracy of a clustering solution as defined in (4). Clustering solutions are plotted in a GA plot based on their granularity and accuracy. Lines are drawn between clustering solutions obtained using the same relatedness measure but different values of the resolution parameter $\gamma$. We use a logarithmic scale for both the horizontal and the vertical axis in a GA plot.

In the interpretation of a GA plot, one should be aware that for any relatedness measure an increase in granularity will always cause a decrease in accuracy. This is a mathematical necessity in our evaluation framework, and therefore it is not something one should be concerned about. A GA plot can be interpreted by comparing the accuracy of different relatedness measures at a specific level of granularity. As explained above, clustering solutions obtained using different relatedness measures normally do not have exactly the same granularity. However, in a GA plot, lines are drawn between different clustering solutions obtained using the same relatedness measure, providing interpolations between these solutions. Based on such interpolations, the accuracy of different relatedness measures can be compared at a specific level of granularity. These comparisons can be performed at different levels of granularity. Sometimes different levels of granularity will yield inconsistent results, with for instance relatedness measure $A$ outperforming relatedness measure $B$ at one level of granularity and the opposite outcome at another level of granularity. In other cases, consistent results will be obtained at all levels of granularity. For instance, relatedness measure $C$ may consistently outperform relatedness measure $D$, regardless of the level of granularity.

In the next two subsections, GA plots will be used to compare different citation-based and text-based relatedness measures.

**4.4. Comparison of citation-based relatedness measures**

For each of our three fields (i.e., cell biology, condensed matter physics, and economics), Figure 1 presents a GA plot for comparing the DC, BC, CC, DC-BC-CC, and EDC citation-based relatedness measures discussed in Subsection 3.1. In the case of the DC-BC-CC relatedness measure, two values of the parameter $\alpha$ are considered, $\alpha = 1$ and $\alpha = 5$. The BM25 text-based relatedness measure discussed in Subsection 3.2 is used as the evaluation criterion. Results obtained when this relatedness measure is used to cluster publications are also included in the GA plots. These results provide an upper bound for the results that can be obtained using the citation-based relatedness measures. (Recall from Subsection 2.3 that the highest possible accuracy is obtained when publications are clustered based on the same relatedness measure that is also used as the evaluation criterion.) All relatedness measures (except for DC and EDC; see Subsection 3.3) use a value of $20$ for the parameter $M$ of the top $M$ relatedness approach.

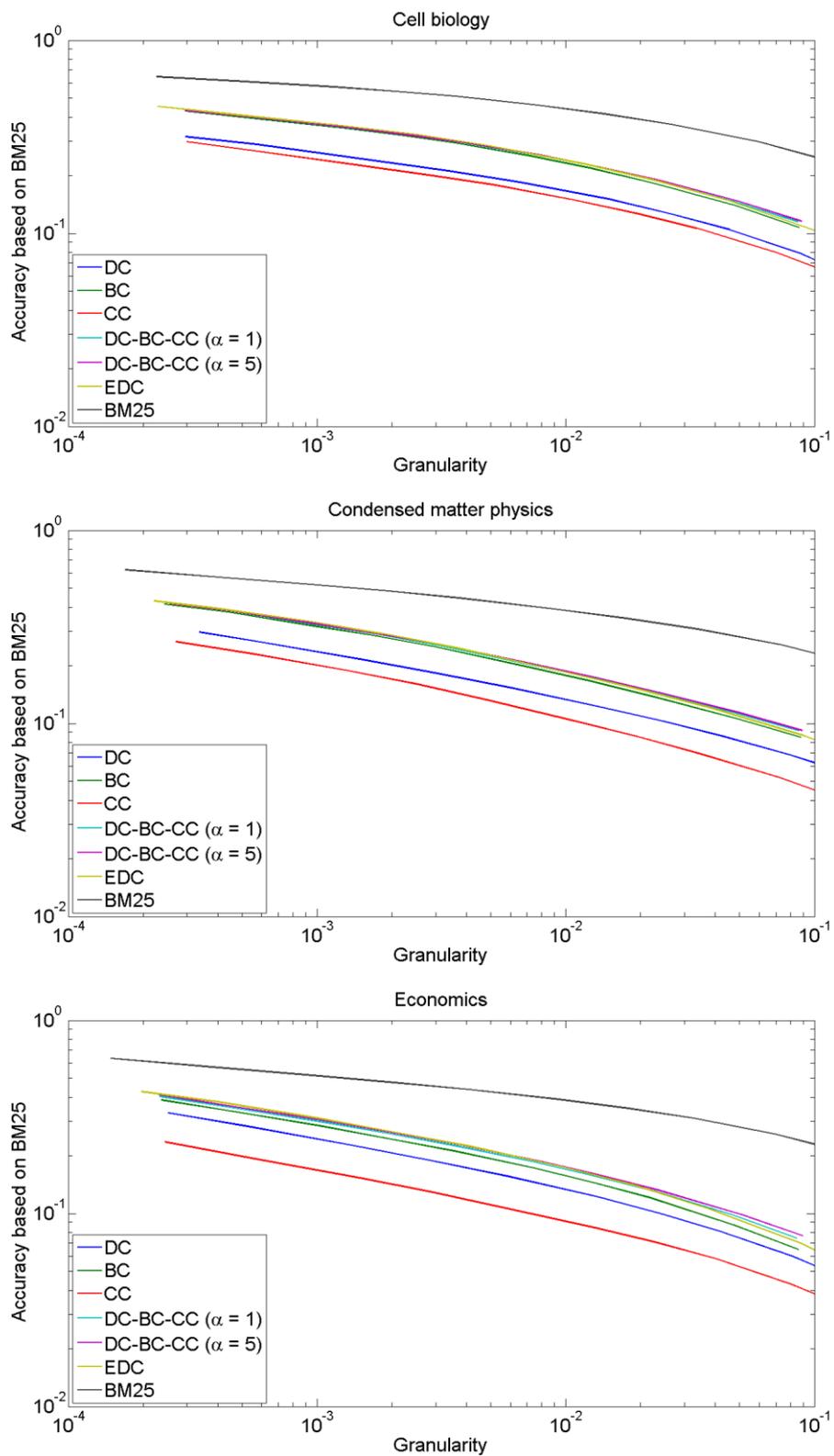

Figure 1. GA plots for comparing citation-based relatedness measures. The BM25 text-based relatedness measure is used as the evaluation criterion.

To interpret the GA plots in Figure 1, it is important to have some understanding of the meaning of the different levels of granularity. For each of our three fields, a clustering solution consists of several hundreds of significant clusters when the granularity is around 0.001, where we define a significant cluster as a cluster that includes at least ten publications. A granularity around 0.01 corresponds with several thousands of significant clusters.

As can be seen in Figure 1, the results obtained for cell biology, condensed matter physics, and economics are fairly similar. Using BM25 as the evaluation criterion, CC has the worst performance of all citation-based relatedness measures. This is not surprising. Uncited publications have no co-citation relations with other publications and therefore cannot be properly clustered. Table 1 shows that in all three fields the percentage of publications without co-citation relations is quite high. This is an important explanation of the bad performance of CC. The bad performance of CC is in line with recent results of Klavans and Boyack (2017). DC outperforms CC but is outperformed by all other citation-based relatedness measures. The performance of DC is especially weak in cell biology. The disappointing performance of DC in all three fields is an important finding, in particular given the increasing popularity of DC in recent years. BC, DC-BC-CC, and EDC all perform about equally well. DC-BC-CC and EDC seem to slightly outperform BC, but the difference is tiny, especially in cell biology and condensed matter physics. Likewise, there is hardly any difference between the parameter values $\alpha = 1$ and $\alpha = 5$ for DC-BC-CC. Our finding that BC and EDC perform about equally well differs from results of Klavans and Boyack, who found that an approach similar to EDC significantly outperforms BC. Our results are based on a more principled evaluation framework and a different evaluation criterion than the results of Klavans and Boyack, which most likely explains why our findings are different from theirs.

To test the sensitivity of our results to the value of the parameter $M$ of the top $M$ relatedness approach, Figure 2 presents a GA plot in which the DC-BC-CC citation-based relatedness measure (with $\alpha = 1$) is compared for different values of $M$. The BM25 text-based relatedness measure is again used as the evaluation criterion. Only the field of condensed matter physics is considered. As can be seen in Figure 2, our results are rather insensitive to the value of $M$.

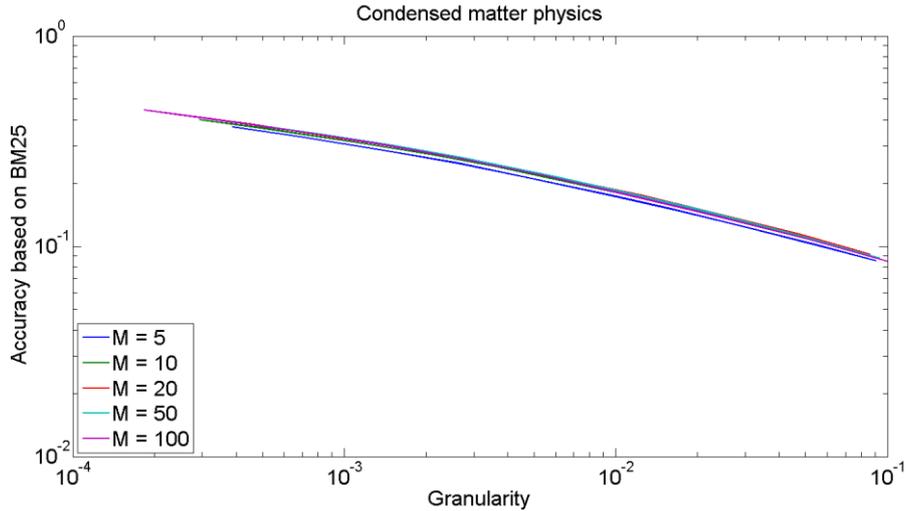

Figure 2. GA plot for comparing the DC-BC-CC citation-based relatedness measure (with $\alpha = 1$) for different values of the parameter $M$ of the top $M$ relatedness approach. The BM25 text-based relatedness measure is used as the evaluation criterion.

We also tested the sensitivity of our results to the choice of the text-based relatedness measure that is used as the evaluation criterion. The results turned out to be insensitive to this choice. Replacing BM25 by CT (with $\beta = 0.5$) yielded very similar results (not shown).

**4.5. Comparison of text-based relatedness measures**

Figure 3 presents GA plots for comparing the BM25 and CT text-based relatedness measures discussed in Subsection 3.2. In the case of the CT relatedness measure, three values of the parameter $\beta$ are considered, $\beta = 0.0$, $\beta = 0.5$, and $\beta = 1.0$. The DC-BC-CC citation-based relatedness measure discussed in Subsection 3.1 (with $\alpha = 1$) is used as the evaluation criterion. Results obtained when this relatedness measure is used to cluster publications are also included in the GA plots. These results provide an upper bound for the results that can be obtained using the text-based relatedness measures. All relatedness measures use a value of 20 for the parameter $M$ of the top $M$ relatedness approach.

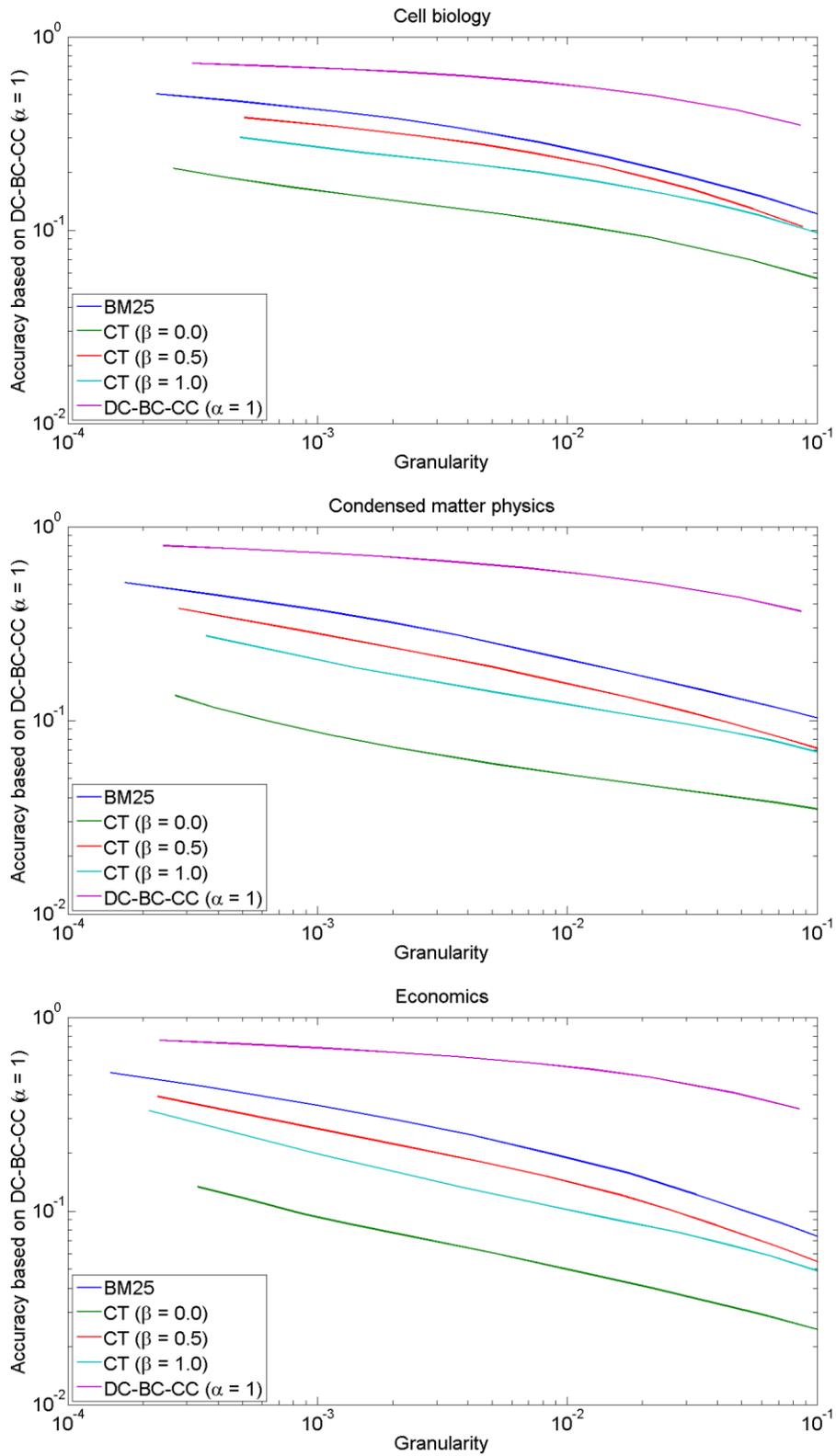

Figure 3. GA plots for comparing text-based relatedness measures. The DC-BC-CC citation-based relatedness measure (with $\alpha = 1$) is used as the evaluation criterion.

The results presented in Figure 3 for cell biology, condensed matter physics, and economics are very similar. Using DC-BC-CC as the evaluation criterion, BM25 outperforms CT, regardless of the value of the parameter $\beta$. The good performance of BM25 is in agreement with the results of Boyack et al. (2011). By far the worst performance is obtained when CT is used with the parameter value $\beta = 0.0$. This confirms the importance of reducing the influence of frequently occurring terms. However, CT with the parameter value $\beta = 0.5$ outperforms CT with the parameter value $\beta = 1.0$. Hence, the influence of frequently occurring terms should not be reduced too strongly.

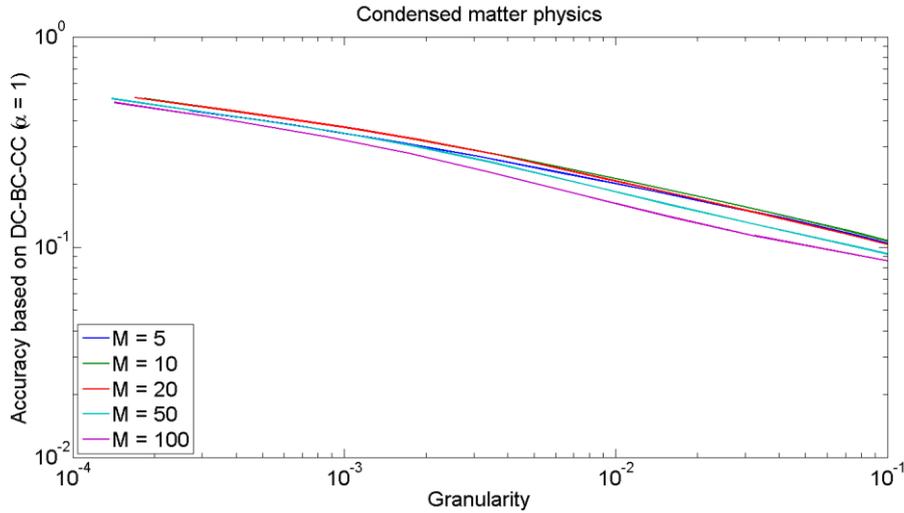

Figure 4. GA plot for comparing the BM25 text-based relatedness measure for different values of the parameter $M$ of the top $M$ relatedness approach. The DC-BC-CC citation-based relatedness measure (with $\alpha = 1$) is used as the evaluation criterion.

To test the sensitivity of our results to the value of the parameter $M$ of the top $M$ relatedness approach, Figure 4 presents a GA plot in which the BM25 text-based relatedness measure is compared for different values of $M$, using the DC-BC-CC citation-based relatedness measure (with $\alpha = 1$) as the evaluation criterion. Only the field of condensed matter physics is considered. Interestingly, and perhaps surprisingly, the highest values of $M$ (i.e., $M = 50$ and $M = 100$) are outperformed by lower values of $M$. Hence, while the highest values of $M$ require most computing time and most computer memory, they yield the lowest accuracy. The highest accuracy is obtained for $M = 10$ or $M = 20$. In line with the approach taken by Boyack et al. (2011), it therefore seems sufficient to keep only the 10 or 20 strongest relations per publication.

We also tested the sensitivity of our results to the choice of the citation-based relatedness measure that is used as the evaluation criterion. The results turned out to be insensitive to this choice. Replacing DC-BC-CC (with $\alpha = 1$) by CC yielded very similar results (not shown).

## 5. Conclusions

The problem of clustering scientific publications involves significant conceptual and methodological challenges. We have introduced a principled methodology for evaluating the accuracy of clustering solutions obtained using different relatedness measures. Our methodology can be applied to evaluate the accuracy of clustering solutions obtained using two relatedness measures $A$ and $B$, where a third relatedness measure $C$ is used as the evaluation criterion. Preferably, relatedness measure $C$ should be as independent as possible from relatedness measures $A$ and $B$. Relatedness measures $A$ and $B$ for instance may be citation-based relatedness measures, and relatedness measure $C$ may be a text-based relatedness measure (or the other way around).

The empirical results that we have presented are based on a large-scale analysis of publications in the fields of cell biology, condensed matter physics, and economics indexed in the Web of Science database. We have used our proposed methodology, complemented with a graphical approach based on so-called GA plots, to compare different citation-based relatedness measures that can be used to cluster publications. Using the BM25 text-based relatedness measure as the evaluation criterion, we have found that co-citation relations and direct citation relations yield less accurate clustering solutions than a number of other citation-based relatedness measures. Bibliographic coupling relations, possibly combined with direct citation relations and co-citation relations, can be used to obtain more accurate clustering solutions. The so-called extended direct citation approach yields clustering solutions with an accuracy that is similar to or even somewhat higher than the accuracy of clustering solutions obtained using bibliographic coupling relations. We note that our analyses have been restricted to individual fields of science. In an analysis that covers all fields of science and a long period of time, differences between the ordinary direct citation approach and the extended direct citation approach can be expected to be much smaller. We have also compared different text-based relatedness measures using a citation-based relatedness measure (obtained by combining direct citation relations, bibliographic coupling relations, and co-citation relations) as the evaluation criterion. BM25 has turned out to yield more accurate clustering solutions than the other text-based relatedness measures that we have studied.

We have also analyzed the use of the so-called top $M$ relatedness approach. This approach can be used to reduce the amount of computing time and computer memory needed to cluster publications. We have found that the use of the top $M$ relatedness approach does not decrease the accuracy of clustering solutions. In fact, in the case of text-based relatedness measures, the accuracy of clustering solutions may even increase.

In this paper, we have adopted the perspective that it is useful to assume the existence of an absolute notion of accuracy. Given the lack of a ground truth, the accuracy of a clustering solution cannot be directly measured. However, by assuming the existence of an absolute notion of accuracy, our methodology allows the accuracy of a clustering solution to be evaluated in an indirect way. An alternative perspective is that there is no absolute notion of accuracy and that it is not meaningful to ask whether one clustering solution is more accurate than another one (e.g., Gläser et al., 2017). From this perspective, clustering solutions obtained using different relatedness measures each provide a legitimate viewpoint on the organization of the scientific literature. We fully acknowledge the value of this alternative perspective, and we recognize the need to better understand how clustering solutions obtained using different relatedness measures offer complementary viewpoints. Nevertheless, from an applied point of view focused on practical applications, we believe that there is a need to evaluate the accuracy of clustering solutions obtained using different relatedness measures and to identify the relatedness measures that yield the most accurate clustering solutions. This motivates our choice to make the assumption of the existence of an absolute notion of accuracy. For those who consider this assumption to be problematic, we would like to suggest that the results provided by our methodology could be given an alternative interpretation that does not depend on this assumption. Instead of interpreting the results in terms of accuracy, they could be interpreted in terms of the degree to which different relatedness measures yield similar clustering solutions.

The most obvious direction for future research is to apply our methodology to a broader set of relatedness measures. Examples include relatedness measures based on full-text data, grant data, and keyword data (e.g., MeSH terms). Some of this work is already ongoing (Boyack & Klavans, 2018).

## Acknowledgements

We would like to thank Dick Klavans, Vincent Traag, and two reviewers for their helpful comments on our work. Part of this research was conducted when Giovanni Colavizza was affiliated with the Digital Humanities Laboratory, École Polytechnique Fédérale de Lausanne, Switzerland. Giovanni Colavizza was in part supported by a Swiss National Fund grant (number P1ELP2_168489).

## Competing interests

The authors use clustering approaches similar to the ones discussed in this paper in commercial applications.

## Appendix A: Motivation for the evaluation framework

In this appendix, we present a conceptual motivation for the framework introduced in Section 2 for evaluating the accuracy of clustering solutions obtained using different relatedness measures. The motivation is based on an analogy with the evaluation of the accuracy of different indicators that provide estimates of values drawn from a probability distribution. We use this analogy because the evaluation of the accuracy of different indicators can be analyzed in a more precise way than the evaluation of the accuracy of clustering solutions obtained using different relatedness measures.

### A.1. Evaluating two indicators using a third indicator

Suppose $N$ values $v_1, \ldots, v_N$ have been drawn from a standard normal distribution. These values cannot be observed directly. However, we have available three indicators $A$, $B$, and $C$ that provide estimates of the values $v_1, \ldots, v_N$. Let the estimates provided by the indicators $A$, $B$, and $C$ be denoted by $v_1^A, \ldots, v_N^A$, $v_1^B, \ldots, v_N^B$, and $v_1^C, \ldots, v_N^C$, respectively. Suppose we need to choose between the use of indicator $A$ or indicator $B$. We therefore want to know which of these two indicators is more accurate. Since the values $v_1, \ldots, v_N$ cannot be observed directly, we cannot evaluate the accuracy of indicators $A$ and $B$ by comparing the estimates $v_1^A, \ldots, v_N^A$ and $v_1^B, \ldots, v_N^B$ with the true values $v_1, \ldots, v_N$. However, if indicator $C$ can be assumed to be independent of indicators $A$ and $B$ (see Appendix A.2 for a further discussion of this assumption), it is possible to use indicator $C$ to evaluate the accuracy of indicators $A$ and $B$. This can be seen as follows.

Suppose the estimates provided by indicators $A$, $B$, and $C$ are given by

$$v_i^A = \sqrt{a_A} v_i + \sqrt{1 - a_A} e_i^A, \tag{A1}$$

$$v_i^B = \sqrt{a_B}v_i + \sqrt{1-a_B}e_i^B, \qquad (A2)$$

$$v_i^C = \sqrt{a_C}v_i + \sqrt{1-a_C}e_i^C, \qquad (A3)$$

where $a_A, a_B, a_C \in [0,1]$ denote the accuracy of indicators $A$, $B$, and $C$ and where $e_i^A$, $e_i^B$, and $e_i^C$ have been independently drawn from a standard normal distribution. Eqs. (A1), (A2), and (A3) imply that the estimates provided by indicators $A$, $B$, and $C$ follow a standard normal distribution. Because $e_i^A$, $e_i^B$, and $e_i^C$ have been independently drawn, we say that indicators $A$, $B$, and $C$ are independent of each other.

We want to know whether $a_A > a_B$ or $a_A < a_B$. To determine this, we calculate the mean squared difference between the estimates provided by indicators $A$ and $C$ and between the estimates provided by indicators $B$ and $C$. This yields

$$\text{MSD}_{AC} = \frac{1}{N}\sum_i(v_i^A - v_i^C)^2, \qquad (A4)$$

$$\text{MSD}_{BC} = \frac{1}{N}\sum_i(v_i^B - v_i^C)^2. \qquad (A5)$$

If $N$ is infinitely large, standard results from probability theory can be used to show that

$$\text{MSD}_{AC} = 2 - 2\sqrt{a_A a_C}, \qquad (A6)$$

$$\text{MSD}_{BC} = 2 - 2\sqrt{a_B a_C}. \qquad (A7)$$

Based on (A6) and (A7), if $\text{MSD}_{AC} < \text{MSD}_{BC}$, then $a_A > a_B$. Conversely, if $\text{MSD}_{AC} > \text{MSD}_{BC}$, then $a_A < a_B$. This shows that indicator $C$ can be used to evaluate the accuracy of indicators $A$ and $B$ and to determine which of the two indicators is more accurate. Moreover, this is possible even if indicator $C$ itself has a low (but non-zero) accuracy, perhaps much lower than the accuracy of indicators $A$ and $B$.

We have now demonstrated how an indicator $C$ can be used to evaluate the accuracy of indicators $A$ and $B$. The idea of the evaluation framework presented in Section 2 is similar, but instead of indicators we consider relatedness measures and clustering solutions obtained using these relatedness measures. We use a relatedness measure $C$ to evaluate the accuracy of clustering solutions obtained using relatedness measures $A$ and $B$. Relatedness measures $A$ and $B$ for instance could be two citation-based measures, such as a measure based on direct citation relations and a measure based on bibliographic coupling relations, while relatedness measure $C$ could be a text-based measure, such as a measure based on BM25. If relatedness measure $C$ can be assumed to be (approximately) independent of relatedness measures $A$ and $B$, it can be used to evaluate the accuracy of clustering solutions obtained using relatedness measures $A$ and $B$. This is possible even if relatedness measure $C$ itself has a lower accuracy than relatedness measures $A$ and $B$.

**A.2. Independence assumption**

In Appendix A.1, we relied on the assumption that indicator $C$ is independent of indicators $A$ and $B$. We now demonstrate the importance of this assumption. To do so, we drop the assumption and we allow for a dependence between indicators $A$ and $C$. Rather than by (A3), suppose estimates provided by indicator $C$ are given by

$$v_i^C = \sqrt{(1-d_{AC})a_C + d_{AC}a_A}\, v_i + \sqrt{(1-d_{AC})(1-a_C)}\, e_i^C + \sqrt{d_{AC}(1-a_A)}\, e_i^A, \tag{A8}$$

where $d_{AC} \in [0,1]$ denotes the dependence between indicators $A$ and $C$. If $d_{AC} = 0$, there is no dependence between indicators $A$ and $C$ and (A8) reduces to (A3). On the other hand, if $d_{AC} = 1$, there is a full dependence between indicators $A$ and $C$. The indicators then provide identical estimates, and (A8) therefore reduces to (A1). Eq. (A8) implies that the estimates provided by indicator $C$ follow a standard normal distribution and that there is no dependence between indicators $B$ and $C$.

Based on (A1), (A2), and (A8), it can be shown that

$$\text{MSD}_{AC} = 2 - 2\sqrt{a_A a_{AC}} - 2\sqrt{d_{AC}(1-a_A)}, \tag{A9}$$

$$\text{MSD}_{BC} = 2 - 2\sqrt{a_B a_{AC}}, \tag{A10}$$

where

$$a_{AC} = (1-d_{AC})a_C + d_{AC}a_A. \tag{A11}$$

As expected, if $d_{AC} = 0$, (A9) and (A10) reduce to (A6) and (A7). It follows from (A9) and (A10) that $\text{MSD}_{BC} < \text{MSD}_{AC}$ if and only if

$$a_B > a_A + \frac{d_{AC}}{a_{AC}}(1 - a_A)^2 + 2\sqrt{\frac{d_{AC}}{a_{AC}}}\sqrt{a_A}(1 - a_A). \tag{A12}$$

If $d_{AC} > 0$ and $a_A < 1$, the sum of the second and the third term in the right-hand side of (A12) is positive. It is then possible that the inequality in (A12) is not satisfied even though $a_B > a_A$. Hence, it is possible that $\text{MSD}_{BC} > \text{MSD}_{AC}$ even though $a_B > a_A$. Indicator $C$ then gives the incorrect impression that indicator $A$ is more accurate than indicator $B$. This is due to the dependence between indicators $A$ and $C$. The higher the dependence $d_{AC}$, the more likely indicator $C$ is to give the incorrect impression that indicator $A$ is more accurate than indicator $B$. In the extreme case in which $d_{AC} = 1$, it is even impossible for indicator $B$ to be considered more accurate than indicator $A$.

We have now demonstrated the importance of the independence assumption when an indicator $C$ is used to evaluate the accuracy of indicators $A$ and $B$. In the evaluation framework presented in Section 2, the independence assumption has a similar importance. When a relatedness measure $C$ is used to evaluate the accuracy of clustering solutions obtained using relatedness measures $A$ and $B$, it is important that relatedness measure $C$ is (approximately) independent of relatedness measures $A$ and $B$. For instance, if there is a dependence between relatedness measures $A$ and $C$, evaluations performed using relatedness measure $C$ will be biased in favor of clustering solutions obtained using relatedness measure $A$.

## Appendix B: Consistent and inconsistent evaluation frameworks

In this appendix, we formally show the consistency of the evaluation framework proposed in Section 2. We also present an example of an inconsistent evaluation framework.

### B.1. Consistency of the proposed evaluation framework

Consider two relatedness measures $X$ and $Y$. Suppose that we have obtained a clustering solution $c_1^X, \ldots, c_N^X$ for relatedness measure $X$ by maximizing the quality function in (2) using the resolution parameter $\gamma^X$. In addition, we have obtained a clustering solution $c_1^Y, \ldots, c_N^Y$ for relatedness measure $Y$ by maximizing the same quality function using the resolution parameter $\gamma^Y$. Suppose also that the two clustering solutions satisfy the condition in (5). Hence, the two clustering solutions have the same granularity. When the accuracy of the two clustering solutions is evaluated using relatedness measure $X$, it is guaranteed that the clustering solution obtained using relatedness measure $X$ will be more accurate than the clustering solution obtained using relatedness measure $Y$. More precisely, it is guaranteed that $A^{X|X} \geq A^{Y|X}$, where $A^{X|X}$ and $A^{Y|X}$ denote the accuracy of the two clustering solutions according to the accuracy measure in (4). This result shows the consistency of our evaluation framework.

To prove the above result, suppose that $A^{X|X} < A^{Y|X}$. It then follows from (4) that

$$\sum_{i,j} I(c_i^X = c_j^X) r_{ij}^X < \sum_{i,j} I(c_i^Y = c_j^Y) r_{ij}^X. \tag{B1}$$

The granularity condition in (5) states that

$$\sum_k (s_k^X)^2 = \sum_l (s_l^Y)^2. \tag{B2}$$

Eqs. (B1) and (B2) imply that

$$\sum_{i,j} I(c_i^X = c_j^X) r_{ij}^X - \gamma^X \sum_k (s_k^X)^2 < \sum_{i,j} I(c_i^Y = c_j^Y) r_{ij}^X - \gamma^X \sum_l (s_l^Y)^2. \tag{B3}$$

It now follows from (2) and (B3) that $c_1^Y, \ldots, c_N^Y$ offers a higher quality clustering solution for relatedness measure $X$ and resolution parameter $\gamma^X$ than $c_1^X, \ldots, c_N^X$. However, this is not possible, since $c_1^X, \ldots, c_N^X$ is defined as the clustering solution that maximizes (2) for relatedness measure $X$ and resolution parameter $\gamma^X$. We therefore have a contradiction. This proves that $A^{X|X} \geq A^{Y|X}$.

A minor qualification needs to be made. In practice, heuristic algorithms are usually used to maximize the quality function in (2). There is no guarantee that these algorithms are able to find the global maximum of the quality function (see

Subsection 4.2). In exceptional cases, this might cause the consistency of our evaluation framework to be violated.

**B.2. Example of an inconsistent evaluation framework**

Consider an evaluation framework in which clustering solutions are compared using (4) subject to a granularity condition requiring that clustering solutions consist of the same number of clusters. This granularity condition, which was used by Klavans and Boyack (2017), replaces the granularity condition in (5). The following example shows that this evaluation framework is inconsistent.

Suppose we have six publications, labeled P1 to P6. Consider two relatedness measures $X$ and $Y$. Tables B.1 and B.2 show the relatedness of the six publications according to relatedness measures $X$ and $Y$, respectively. Suppose that the resolution parameter $\gamma$ is set to a value of 1.1. For relatedness measure $X$, maximization of the quality function in (2) then yields two clusters, one consisting of publications P1 to P3 and the other consisting of publications P4 to P6. For relatedness measure $Y$, we also obtain two clusters, one consisting of publications P1 to P5 and the other consisting only of publication P6. Since the two clustering solutions both consist of two clusters, our granularity condition is satisfied.

Table B.1. Relatedness of publications according to relatedness measure $X$.

|    | P1 | P2 | P3 | P4 | P5 | P6 |
|----|----|----|----|----|----|----|
| P1 |    | 2  | 2  | 1  | 1  | 1  |
| P2 | 2  |    | 2  | 1  | 1  | 1  |
| P3 | 2  | 2  |    | 1  | 1  | 1  |
| P4 | 1  | 1  | 1  |    | 2  | 2  |
| P5 | 1  | 1  | 1  | 2  |    | 2  |
| P6 | 1  | 1  | 1  | 2  | 2  |    |

Table B.2. Relatedness of publications according to relatedness measure $Y$.

|    | P1 | P2 | P3 | P4 | P5 | P6 |
|----|----|----|----|----|----|----|
| P1 |    | 2  | 2  | 2  | 2  | 1  |
| P2 | 2  |    | 2  | 2  | 2  | 1  |
| P3 | 2  | 2  |    | 2  | 2  | 1  |
| P4 | 2  | 2  | 2  |    | 2  | 1  |
| P5 | 2  | 2  | 2  | 2  |    | 1  |
| P6 | 1  | 1  | 1  | 1  | 1  |    |

Based on (4), we now compare the two clustering solutions. Using relatedness measure $Y$ to evaluate the accuracy of the clustering solutions, we obtain $A^{X|Y} = 10/3$ and $A^{Y|Y} = 20/3$. Hence, as we would intuitively expect, according to relatedness measure $Y$, the clustering solution obtained using relatedness measure $Y$ is more accurate than the one obtained using relatedness measure $X$. Let us now use relatedness measure $X$ to evaluate the accuracy of the clustering solutions. This yields $A^{X|X} = 4$ and $A^{Y|X} = 28/6$. In other words, we obtain the counterintuitive result that, according to relatedness measure $X$, the clustering solution obtained using relatedness measure $X$ is less accurate than the one obtained using relatedness measure $Y$. This shows the inconsistency of our evaluation framework.

## Appendix C: Extended direct citation approach

In this appendix, we discuss the technical details of the extended direct citation approach introduced in Subsection 3.1.

Our aim is to cluster publications $1, \ldots, N$. We refer to these publications as our focal publications. To cluster the focal publications, we also consider publications $N + 1, \ldots, N^{\text{EXT}}$. Each of these non-focal publications has a direct citation relation with at least two focal publications. For $i = 1, \ldots, N$ and $j = 1, \ldots, N^{\text{EXT}}$, the relatedness of publications $i$ and $j$ in the extended direct citation approach is given by

$$r_{ij} = \max(c_{ij}, c_{ji}), \tag{C1}$$

where $c_{ij}$ indicates whether publication $i$ cites publication $j$ ($c_{ij} = 1$) or not ($c_{ij} = 0$).

Following the ideas presented in Subsection 3.4, for $i = 1, \ldots, N$ and $j = 1, \ldots, N^{\text{EXT}}$, the normalized relatedness of publication $i$ with publication $j$ in the extended direct citation approach equals

$$\hat{r}_{ij} = \frac{r_{ij}}{\sum_k r_{ik}}. \tag{C2}$$

To accommodate the non-focal publications $N+1, \ldots, N^{\text{EXT}}$, the quality function in (1) needs to be adjusted. In the extended direct citation approach, publications $1, \ldots, N^{\text{EXT}}$ are assigned to clusters $c_1, \ldots, c_{N^{\text{EXT}}}$ by maximizing the quality function

$$Q = \sum_{i=1}^{N}\left[\sum_{j=1}^{N} I(c_i = c_j)(\hat{r}_{ij} - \gamma) + \sum_{j=N+1}^{N^{\text{EXT}}} I(c_i = c_j)\hat{r}_{ij}\right]. \quad \text{(C3)}$$

The non-focal publications are treated in a special way in (C3). The costs and benefits of assigning a publication to a cluster are different for the non-focal publications than for the focal ones. On the one hand, there is no cost in assigning a non-focal publication to a cluster. To see this, notice that there is no subtraction of $\gamma$ in the second term within the square brackets in (C3). On the other hand, non-focal publications do not yield benefits in the same way as focal publications do. To see this, notice that the outer summation in (C3) extends only over the focal publications. The non-focal publications are not included in this summation.

After the quality function in (C3) has been maximized, we discard the cluster assignments $c_{N+1}, \ldots, c_{N^{\text{EXT}}}$ of the non-focal publications, since we are interested only in the cluster assignments $c_1, \ldots, c_N$ of the focal publications.